\documentclass{article}

\usepackage{arxiv}
\usepackage[utf8]{inputenc} 
\usepackage[T1]{fontenc}    
\usepackage{hyperref}       
\usepackage{url}            
\usepackage{booktabs}       
\usepackage{amsfonts}       
\usepackage{nicefrac}       
\usepackage{microtype}      
\usepackage{graphicx}

\title{Gasp: A DeFi Application Specific Rollup as a Consolidation Layer for All Assets}

\author{
  Stanislav Vozarik \quad
  Mateusz Nowakowski \quad
  Shoeb Siddiqui \quad
  Peter Kris \quad
  Elliot Hill \quad
  Gleb Urvanov
}

\date{}

\begin{document}
\maketitle

\begin{abstract}
Gasp is a decentralized exchange designed as an application-specific Layer 2 (L2) rollup with omnichain connectivity, leveraging EigenLayer's restaked ETH for computation correctness and finalization. With a goal of being a consolidation layer for all crypto assets, the Gasp platform employs optimistic rollup technology to facilitate gas-free, native cross-chain swaps without reliance on traditional bridges, ensuring tokens retain their original L1 grade security. By combining an app-chain architecture with escape hatch mechanisms, Gasp guarantees withdrawal, while MEV minimization through Themis architecture reduces value extraction risks. Gasp's proof-of-liquidity framework unlocks staked liquidity, enhancing capital efficiency and liquidity depth by integrating staking with liquidity provisioning. Additionally, the protocol introduces a time-based reward mechanism, incentivizing long-term liquidity commitment via an asymptotic reward curve. This paper examines the current challenges in cross-chain communication, delineates Gasp's architectural innovations and security guarantees, and examines novel approaches to optimizing DeFi ecosystems.
\end{abstract}


\section{Introduction}
The stratospheric growth of the cryptoasset economy during the last decade has introduced innovation on a scale seldom seen in other industries; but it has also given rise to a multitude of diverse and fragmented mutually independent ecosystems of value \cite{augusto2022multi}. Stemming from just a handful of Layer-1 (L1) blockchain environments such as Bitcoin and Ethereum, these ecosystems now encompass a plethora of networks, each with unique architectures, consensus mechanisms, and security models \cite{augusto2022multi, zhu2023study, augusto2024sok}. While this diversity has spurred innovation, it has also introduced significant challenges to interoperability, particularly the efficient and secure movement of assets, value, and capital across disparate blockchain ecosystems.

Traditional solutions, such as bridges and custodial intermediaries, often fail to address the core issues of trust, security, capital efficiency, and most importantly decentralization, leading to fragmented liquidity and limited usability for decentralized finance (DeFi) applications. Despite the advancements in L2 rollups as scaling solutions and their ability to handle basic token bridging, transferring value between sovereign L1 blockchains remains a significant—yet previously unsolved—challenge. Existing cross-chain solutions, including centralized exchanges (CEXes) and token bridges, introduce trust and security risks that many consider untenable for the long-term growth of DeFi.

Centralized exchanges remain the dominant method for cross-chain value transfer, but their reliance on custodianship and trust contradicts the basic principles of decentralization, as well as contradicting a general user trend towards decentralized infrastructure adoption. The core issue with centralized exchanges, that of trust in a centralized entity, was highlighted in the most high-profile of scandals through the downfall of FTX, which led to losses in the billions of dollars for users and the crypto industry as a whole. Though, centralized risk and mismanagement aside, there are other more practical limitations to the CEX experience for users.

Users transferring assets cross-chain using a general CEX interface, for instance from Ethereum to Polygon, must first deposit tokens to a CEX or acquire crypto through fiat deposits, approve the assets for trading, identify a pair with sufficient liquidity and execute a trade, move assets from their trading wallet to their funding wallet, and withdraw to the destination chain—an approach that sacrifices security and user autonomy.

Additionally, centralized exchanges often impose restrictions, including withdrawal limits and trading suspensions during periods of market volatility, undermining user autonomy. Geographic and regulatory constraints further hinder accessibility, while KYC requirements raise privacy concerns for users. These issues, coupled with liquidity fragmentation across platforms, highlight the growing need for trust-minimized, decentralized solutions that preserve user security and align with the permissionless ethos of blockchain technology.

Token bridges, though promising in theory and widely used by crypto natives, also pose systemic risks \cite{li2022blockchain, zhang2023security}. Bridges, while operating directly on the blockchain and thus to some degree offering a decentralized experience compared to that of a CEX, often employ various and sophisticated multi-signature mechanisms to facilitate cross-chain transfer, the decentralization of which is oftentimes questionable \cite{lee2023sok}. A core limitation of bridges lies in their reliance on bridged tokens. These tokens expose users to risks associated with both the bridge's security and the source blockchain. If the source chain experiences a $51\%$ attack or rollback, the value of bridged assets on the destination chain may be irreparably compromised. Additionally, bridged tokens have new security assumptions of the destination chain. Bridged tokens lack intrinsic utility beyond trading, making them an inefficient and risky solution for interchain value transfer.

As a result, hacks from cross-chain bridges have led to monetary losses of around 2.9 billion USD to date, with a recent systematic review revealing that a substantial portion ($65.8\%$) of these losses originated from projects secured by intermediary permissioned networks with unsecured cryptographic key operations \cite{augusto2024sok}. Bridges magnify risks as their complexity grows with each connected ecosystem, often requiring compromises in decentralization through mechanisms like multi-signature governance, and therefore we believe they are not a scalable and secure solution for the problem of cross-chain communication.

An alternative approach which aims to preserve decentralization and user experience are atomic swaps. First demonstrated between Bitcoin and Litecoin in 2017, atomic swaps enable secure, trustless exchanges of assets across chains without requiring intermediaries or bridged tokens. However, while atomic swaps align with the principles of decentralization and security, they have yet to achieve widespread adoption across the expressive and diverse ecosystems of modern blockchains like Ethereum, Solana, and Cosmos.

Despite their promise, atomic swaps come with several challenges and risks that have hindered their broader adoption. One key limitation is the reliance on advanced scripting capabilities and cross-chain compatibility, which are not uniformly supported across all blockchains. For example, the lack of standardized smart contract functionality on some chains limits the implementation of atomic swaps. Additionally, atomic swaps can suffer from liquidity constraints, as they require counterparties to hold the specific assets being exchanged, reducing the flexibility of trading compared to centralized exchanges or liquidity pools.

The user experience for atomic swaps also presents a multitude of hurdles. The process typically requires multiple onchain transactions, increasing fees and execution times, which can deter adoption, especially on chains with high gas fees. Furthermore, atomic swaps often necessitate complex cryptographic protocols like hash time-locked contracts, making them less accessible to non-technical users. We believe the underlying motivations behind atomic swaps are great guiding principles. They provide the universal accessibility of tokens that are currently attributed to centralized exchanges while benefiting from the security, trustlessness, and decentralization that only distributed ledgers can offer. While promising, we do not believe that atomic swaps can currently address the full scope of challenges needed to solve cross-chain communication, though they remain a promising avenue of research.

As we have discussed above and previous authors have elucidated, current systems often suffer from operational inefficiencies, high transaction costs, and security vulnerabilities, which hinder the seamless integration of decentralized applications and inhibit the realization of a truly interconnected blockchain network. These challenges are particularly pronounced in environments requiring trustless interoperability across chains with varying levels of decentralization and cryptographic assurance, where existing solutions introduce orders of magnitude greater risks when compared to native L1 security. As a solution for these challenges, we introduce Gasp, a consolidation layer for all crypto assets.

Gasp, an application-specific Layer 2 (L2) rollup, addresses these limitations by providing a unified platform for omnichain connectivity. By adopting a consolidation layer approach, Gasp offers compatibility with all blockchains, including EVM-compatible chains, non-EVM ecosystems, and emerging blockchain architectures.

\subsection{Standard Rollup Terminology}
The common notion of one blockchain using the tokens and state of another blockchain is expressed in terms of layers. In conventional blockchain architectures, Layer 2 (L2) solutions enhance scalability by offloading transaction processing from the main Layer 1 (L1) chain. Rollups, a prevalent L2 strategy, bundle multiple transactions into a single batch, executing them offchain and subsequently submitting a concise proof to the L1 chain. This method reduces transaction costs and alleviates computational demands on the L1 network.

The blockchain making use of another blockchain is seen as 'sitting' on top of the underlying chain. A Layer 2 (L2) chain sits on top of a Layer 1 (L1) chain. Most commonly, L2s are built as scaling solutions to make transactions on the L2 cheaper than the same transaction on the L1. One of these mechanisms of compressing the state of the L2 for the L1 is referred to as a rollup.

The nodes that are responsible for block production by sequencing the transactions and executing them to produce state transitions on the L2 are referred to as sequencers \cite{li2023decentralized}. Sequencers play a pivotal role in rollup operations by managing transaction ordering and state updates. However, the reliance on a single, centralized sequencer in many current rollup implementations raises concerns regarding centralization, potentially undermining the decentralized ethos of blockchain systems. Many rollups today still rely on a single sequencer and are thus generally considered to be still centralized. Gasp will use decentralized sequencers by distributing the sequencing function across multiple nodes, this design enhances resilience, reduces the risk of censorship, and aligns with the principles of decentralization.

A core property of blockchains is that they continuously produce blocks - i.e., they maintain liveness. To ensure that users are still able to withdraw assets even if an L2 loses liveness, a fallback mechanism can be put in place that allows withdrawal directly from the L1. This is commonly referred to as an "escape hatch".

\subsection{External State}
Rollups allow us to build and scale to L2s while relying only on the security of the underlying L1. Together they form an ecosystem. A problem occurs when we want to transact with the state and assets originating in other ecosystems like we would do when bridging tokens or doing cross-chain swaps.

Cross-chain interactions introduce significant complexity when managing external states, as they require reconciling data and state across disparate blockchain ecosystems. For example, swapping DOT (Polkadot) for ETH (Ethereum) necessitates not only ensuring the DOT state on Polkadot's blockchain is updated correctly but also synchronizing this change with Ethereum's ledger for ETH. In the case of transactions between ecosystems, communication between them becomes necessary. From the perspective of Ethereum, ownership of DOT is 'external state' information. This external state information originates from another chain, outside of Ethereum, and can thus not be proven by Ethereum.

Rollups, while effective at enhancing scalability and reducing transaction costs, traditionally face limitations in addressing external state issues. Rollups are designed to operate within a single blockchain ecosystem, optimizing transaction throughput and cost-efficiency for the native chain. However, their independence from external blockchains makes cross-chain state reconciliation inherently complex.

A single rollup obtains security from the L1 it is built upon. This has led to the notion that there is never a situation where it makes sense for an L2 to sit on top of multiple L1s, as a rollup cannot obtain additional security from multiple L1s. The argument is that rolling up the same state transition to multiple L1s could lead to situations where different L1s would contradict each other on such a state and the L2 has to defer to the truth of only one L1. Consequently, if the L2 relies on a single L1 in such cases, the other L1s did not add to the rollup's security to begin with. Thus it is commonly assumed that only a single L1 can finalize state transitions of the L2 and verify the correct execution of the L2 rules.

Within this notion there is a hidden assumption that all state on the L2 is a monolithic unit that can only be secured by a single L1. Instead, consider that the L2 state can be seen as a sum of all rollup states. This now opens the possibility to find more nuanced solutions that roll up the different state components of the L2 state into their respective L1s.

To overcome these challenges, solutions like bridge protocols or cross-chain messaging standards are often employed. These mechanisms facilitate the transfer of information and state between blockchains, but they introduce additional layers of complexity and potential vulnerabilities as previously discussed at length.

Gasp’s master-rollup model, secured by EigenLayer as we will discover below, overcomes these limitations by coordinating state transitions across multiple rollups. While a single rollup relies on its L1 for security, the master-rollup manages discrete state components, allowing each state to inherit the security of its respective L1. For example, Ethereum-based tokens on the master-rollup retain Ethereum-grade security.

\section{Gasp's Master-Rollup Interchain Infrastructure}
Gasp’s master-rollup interchain infrastructure enables cross-chain decentralized exchange without reliance on traditional bridging mechanisms. It introduces an application-specific chain designed to operate as an additional layer above multiple blockchains, maintaining full L1 security for all tokens. The architecture features decentralized sequencing, supports cross-chain trading with native assets, and incorporates escape hatches for user safety \cite{vozarik2023cross}.

Gasp is built around core design principles which include:
\begin{enumerate}
    \item The facilitation of seamless deposits and withdrawals.
    \item Maintaining the security guarantees of the underlying L1.
    \item Ensuring user access to funds from connected chains in adverse conditions.
    \item Operating in a fully decentralized manner, free of centralized components or entities.
    \item Avoiding reliance on Turing completeness.
\end{enumerate}
This foundational design enables secure and efficient cross-chain atomic swaps. Gasp’s master-rollup architecture’s design goals, properties, and mechanisms are detailed below.

\subsection{Facilitation of Seamless Deposits and Withdrawals}
Gasp is designed to allow seamless deposits and withdrawals from all connected chains, beginning with EVM chains but later expanding to all blockchain ecosystems (including Bitcoin), and even further, to traditional financial infrastructure.

To enable this, Gasp implements a Master-rollup design to facilitate the seamless deposit and withdrawal of assets to and from the Gasp L2 from a number of external blockchain environments. This involves a primary rollup coordinating multiple subordinate rollups, enhancing scalability and interoperability by aggregating and managing their transactions on a shared layer-1 blockchain. Gasp integrates with multiple ecosystems, leveraging cryptographic proofs to enable cross-chain trading with native assets while maintaining safety mechanisms like escape hatches.

A key mechanism for ensuring seamless deposits and withdrawals is Gasp’s implementation of Rolldowns. Rolldowns prevent incorrect L1 reads from corrupting cross-chain states. Pending reads are queued and verified during a dispute period, ensuring only valid reads are processed. Sequencers stake capital on L1 to disincentivize malicious actions, with slashing penalties for misbehavior. This design ensures that even with a majority of malicious sequencers, a single honest actor can maintain system integrity. Gasp employs a robust cross-chain architecture that leverages the described roll-down mechanism to facilitate seamless deposits and withdrawals across interconnected ecosystems.

\subsection{Maintain the Security Guarantees of the Underlying L1}
Gasp maintains the security guarantees of the underlying L1 for all user-held assets, avoiding the introduction or alteration of security assumptions. This is made possible by enabling cross-chain trading with native assets, enabling users to trade native tokens without the need for bridged assets, such as bridged ETH on Polygon or bridged MATIC on Ethereum, ensuring that withdrawals always yield native assets. Additionally, through L1 Security Guarantees, Gasp maintains the robust security provided by layer-1 blockchains, ensuring decentralized consensus, immutability, and protection against attacks for protocols and applications built on them.

Unlike app chains which rely solely on the value of native tokens for security, the master-rollup ensures tokens inherit the robust security of their originating L1 chain. Execution correctness is reinforced by re-execution chains.

\subsection{Ensure User Access to Funds from Connected Chains in Adverse Conditions}
Ensuring user access to funds from connected chains with smart contract capabilities, even in adverse conditions, is made possible through Gasp’s escape hatch functionality. Escape hatches in crypto are mechanisms allowing users to withdraw funds or interact directly with a protocol's underlying layer in case of failures or emergencies, ensuring security and access continuity. Gasp users retain the ability to withdraw funds directly to the L1 layer under all circumstances, ensuring resilience against system failures.

Escape hatches guarantee users the ability to withdraw funds directly to the L1 if the rollup loses liveness. The mechanism automatically pauses state updates during liveness disruptions and resumes operations once synchronization is restored. This design ensures that users retain access to their assets even in the event of sequencer failures.

\subsection{Operate in a Fully Decentralized Manner, Free of Centralized Components or Entities}
Rolldowns facilitate seamless cross-chain deposits and withdrawals. However, rolldowns may introduce delays to transaction finality. One of Gasp’s key design goals necessitates that the protocol operates in a fully decentralized manner, free of centralized components or entities. Gasp achieves this through a mechanism termed ‘Ferries’ or the ‘Ferry mechanism’. Ferries expedite L1 deposits to the master-rollup by fronting funds for users and reclaiming them post-dispute period. This mechanism reduces delays associated with rolldowns, enhancing the user experience while preserving security.

Similarly, decentralized sequencers order transactions in blockchain networks by distributing the task across multiple nodes, ensuring fairness, resilience, and censorship resistance, while reducing centralization risks in rollups and layer-2 solutions. Gasp launches with decentralized sequencers from inception, avoiding the centralization issues seen in many rollup solutions.

\subsection{Avoid Reliance on Turing Completeness}
Gasp avoids reliance on Turing Completeness by deploying through its own application specific blockchain. An application-specific chain is a blockchain tailored for a single application, optimizing performance, security, and customization for its specific use case, unlike general-purpose blockchains. By hosting a single application—the Gasp decentralized exchange—this architecture avoids the inefficiencies of generalized blockchains, including shared blockspace competition. This specificity allows for optimizations like maximal extractable value (MEV) mitigation, gas-free swaps, and proof-of-liquidity mechanisms.

To validate cross-chain transactions, the runtime block of the master-rollup provides cryptographic proofs of state transitions. Finalization proofs ensure that runtime operations are consistent with the underlying L1 rules. By leveraging Proof of Finality through cryptographic state transition proofs, the need for Turing completeness in cross-chain transactions is significantly reduced because the execution and verification of complex logic are shifted to provably correct, pre-computed finality proofs. Instead of requiring a Turing-complete runtime to validate intricate state transitions directly onchain, the system offloads computational complexity to offchain processes that generate compact, verifiable proofs.

\subsection{Gasp's Solution Cross-L1 information transfer via Master-Rollup secured by Eigenlayer}
Having explored the inherent risks and design challenges associated with traditional rollups and L2's for cross-chain communication, Gasp introduces a novel master-rollup architecture which leverages EigenLayer as a settlement layer to achieve decentralized, secure, and efficient cross-L1 communication.

Gasp introduces decentralized sequencers to mitigate centralization risks and employs novel roll-down mechanics to ensure the correctness of cross-chain state transitions. Sequencers are required to stake collateral on L1, with penalties enforced through slashing mechanisms on EigenLayer for incorrect or malicious behavior, ensuring a high degree of accountability. This design guarantees that even a single honest sequencer can invalidate incorrect reads, replacing the traditional $51\%$ consensus requirement with an "at least one honest actor" paradigm.

For state verification, the architecture incorporates signatures to validate operations on L2, ensuring cryptographic accuracy in all updates reported to L1. The re-execution chain, implemented on EigenLayer, enforces deterministic finality by re-executing and verifying all operations with immutable rules. Finalization is achieved through staker signatures onchain validation.

To address latency in cross-L1 communication, Gasp’s architecture introduces a relay mechanism termed "ferries." These entities accelerate L1-to-L2 information transfers by staking proportional collateral on L2 to guarantee the correctness of their relayed data. Incorrect relays result in forfeited collateral, incentivizing reliable operation while reducing delays for users.

Furthermore, the inclusion of escape hatches ensures resilience against L2 liveliness failures, enabling users to securely withdraw assets under predefined conditions. Additionally, the architecture supports decomposed operations for multichain communication, abstracting complex logic into atomic transfers that can be verified and updated across connected L1s. This enables seamless multichain interoperability.

\section{The Gasp DEX}
This research has culminated in the launch of the Gasp cross-chain DEX, the first DEX of its kind to launch in the EigenLayer ecosystem. Gasp is designed to address the design challenges outlined earlier in this paper, namely seamless transfers from multiple blockchain environments, maintaining the security guarantees of the underlying L1 through restaking, ensuring access to funds even in adverse conditions through our escape hatch functionality, and operating in a decentralized manner.

Unlike traditional solutions, Gasp emphasizes a user-centric approach with features that eliminate complexity and enhance efficiency both for the everyday crypto user, professional traders, and B2B integrations. Users benefit from gas-free transactions, achieved through a unique architecture that avoids double-charging and simplifies cost structures. Trades are facilitated through a $0.3\%$ exchange commission fee, which sustains the network and incentivizes liquidity providers, fostering a healthy trading environment.

Gasp employs decentralized liquidity pools and automated vaults to ensure consistent access to liquidity. Liquidity providers (LPs) earn rewards through time-incentivized liquidity mining, which balances risks like impermanent loss with dynamic fee adjustments. Vaults manage assets securely, allocating liquidity strategically to maximize returns while ensuring transaction finality. The system also includes an innovative Ferry Mechanism, as explored earlier in this text, that expedites fund transfers between chains, reducing delays and slippage for users.

Designed for scalability, Gasp supports high-speed trading for both retail and enterprise users, with functionality optimized for solvers, intent layers, and token holders needing cross-rollup compatibility. By integrating staking requirements via the GASP token, the platform ensures network stability and participant alignment, and ensures spam protection in our gas-free application specific chain. 

In the following sections we will expand upon three core concepts that, in addition to our deep research for cross-chain communication, further improve the user experience, including:
\begin{enumerate}
    \item Themis Architecture for MEV minimisation,
    \item Proof of Liquidity on Gasp,
    \item Time-Based LP incentives.
\end{enumerate}

\subsection{Themis Architecture: Gasp's Solution to Maximum Extractable Value (MEV) Challenges}
Gasp features one of the most advanced MEV minimization mechanisms of any DEX developed to date, which we present under the name 'Themis Architecture' \cite{vozarik2023themis}. Themis architecture is presented to be a solution for MEV on application-specific blockchains. The current iteration of the protocol is focusing on the minimization of MEV by combatting two major methods of value extraction: 1) Value Extraction by Reordering (VER); and 2) Value Extraction by Denial (VED). Each value extraction method and Gasp's solution through Themis Architecture is presented in the paragraphs below.

Value Extraction by Reordering (VER) refers to the process where transaction order within a blockchain or distributed ledger system is manipulated to extract economic value. This typically involves reordering, inserting, or censoring transactions within a block to gain financial advantages, such as frontrunning or sandwich attacks in decentralized finance (DeFi) applications \cite{bartoletti2023theoretical, braga2024redistribution}. VER is particularly associated with entities that have control over transaction ordering, such as miners, validators, or sequencers, or any relayer; and it can compromise the fairness, security, and efficiency of the system. By exploiting the temporal ordering of transactions, VER introduces risks like reduced user trust and increased costs for participants in blockchain ecosystems \cite{vozarik2023themis, bartoletti2023theoretical, braga2024redistribution}.

Gasp's VER solution builds on the assumption that block production consists of two responsibilities: deciding on block contents and deciding on execution order. We propose to split the block production into a two-step process, block building and block execution, wherein the first step transactions are accepted into a block ("block building") and in the second step the transaction execution order is shuffled using information that doesn't exist at the block building time ("block execution"). Every miner has a responsibility to build the new block and execute the previous block.

\begin{figure}[htbp]
    \centering
    \includegraphics[width=0.75\linewidth]{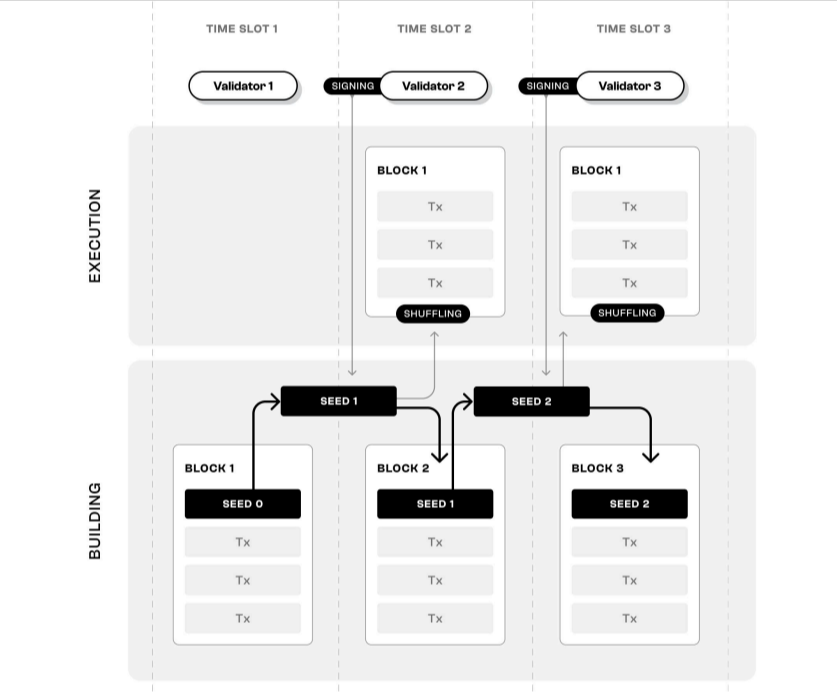}
    \caption{Themis Architecture - The Role of Block Reordering and Timeslots in MEV Minimization.}
    \label{fig:themis_reordering}
\end{figure}

In the first step, shown in Figure \ref{fig:themis_reordering}, the block builder gathers the transactions from the mempool and builds the block. After which, they provide a seed for the subsequent miner that will use it for shuffling the transaction execution order. In the second step, the block executor signs the seed with their private key and uses that as a seed for shuffling. A private key is used because it's unknown to the block builder and can't be manipulated by the block executor. The signature is not malleable and the scheme is deterministic. After that, the executor builds a new block (performing the first step for the next block) and provides the private-key-signed seed.

This effectively creates a seed chain that is used for transaction shuffling. This separation of concerns guarantees that the block builder can't affect the execution order and the block executor can't set the block content and has to shuffle the transactions. In consequence, this creates a 2-block HEAD of the blockchain and doubles the time of the block execution.

Conversely, Value Extraction by Denial (VED) is a mechanism by which economic value is extracted in blockchain systems through the deliberate denial or censorship of specific transactions. This occurs when an entity, such as a miner, validator, sequencer or relayer withholds inclusion of certain transactions in a block, either to favor competing transactions, enforce strategic market outcomes, or extort higher fees from users \cite{vozarik2023themis, bartoletti2023theoretical}. VED exploits the permissionless and decentralized nature of blockchain networks by leveraging control over transaction inclusion, leading to economic inefficiencies, reduced user trust, and potential centralization risks within the system \cite{vozarik2023themis, bartoletti2023theoretical}.

\begin{figure}[htbp]
    \centering
    \includegraphics[width=0.75\linewidth]{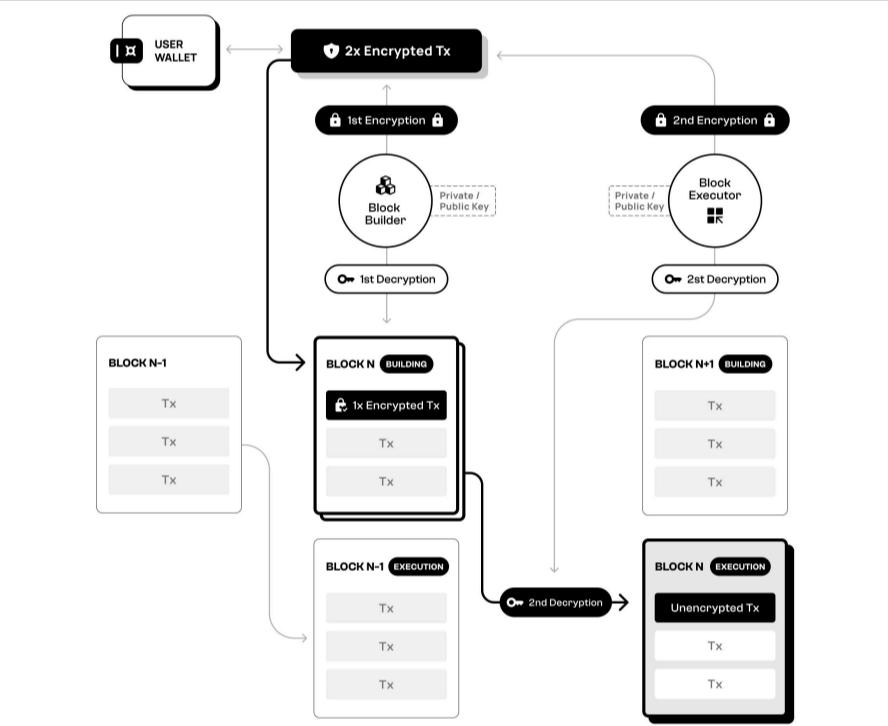}
    \caption{Themis Architecture - Encryption of User Transactions and Block Execution on Gasp.}
    \label{fig:themis_encryption}
\end{figure}

Prevention of VED is made possible by the encryption of transactions by the user as shown in Figure \ref{fig:themis_encryption}, which can be decrypted only by the transaction executor. It is achieved by transactions being encrypted by public keys of both the block builder and block executor. This creates a doubly encrypted transaction that is sent to the mempool.

As a result, the protocol combats value extraction on multiple levels, paving the way for fair ordering and better slippage for the users, in comparison to similar applications where value extraction is sustained by redistribution mechanisms or simply ignored.

\subsection{Proof of Liquidity (PoL)}
Gasp uses Proof-of-Liquidity (PoL), a novel consensus mechanism designed to address the inherent trade-off in decentralized finance (DeFi) ecosystems between network security and liquidity provision \cite{vozarik2023proof}. By enabling liquidity pool (LP) tokens to serve as staking assets, PoL simultaneously secures the network and maintains liquid markets. This mechanism ensures that liquidity provided on Gasp enhances security and liveness of the rollup through being allocated in staking. The mechanism can be considered as a pre-confirmation as well.

The implementation of PoL introduces several implications. First, it establishes a liquidity base layer, as LP tokens paired with the GASP native token receive block rewards, creating a foundational liquidity pool for other trading pairs. Second, PoL combines rewards from staking and liquidity provision, offering dual income streams through block rewards and trading fees, thus incentivizing participation. However, this model also introduces risks, including slashing penalties associated with misbehaving nodes and impermanent loss risk for LPs. These risks are well-documented elsewhere and can be mitigated through governance mechanisms, including the whitelisting of eligible staking assets to prevent gaming of the system.

By unlocking staked liquidity and providing additional revenue streams to participants, PoL enhances the economic dynamics of the Gasp DEX. Its focus on capital efficiency positions it as a transformative approach in the evolution of staking mechanisms, and enables Gasp to optimize liquidity, security, and align user incentives.

\subsection{Time-Based LP Incentives}
\begin{figure}[htbp]
    \centering
    \includegraphics[width=0.85\linewidth]{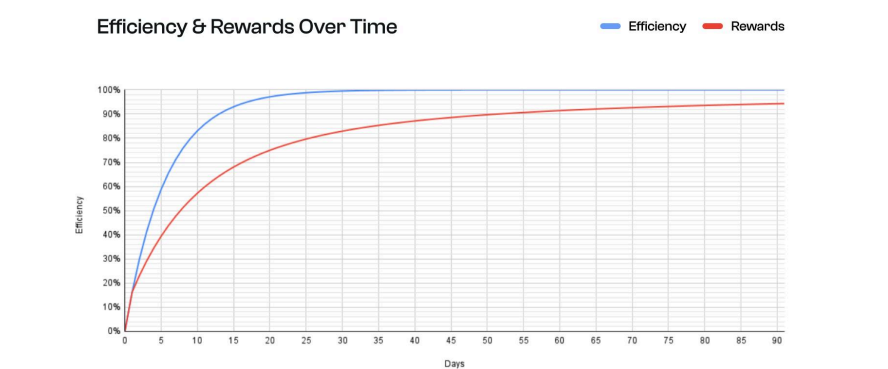}
    \caption{Efficiency \& Rewards Over Time in Gasp's Time-Based LP Incentives.}
    \label{fig:time_incentives}
\end{figure}

Through time-based liquidity provider (LP) incentives, Gasp introduces a novel mechanism \cite{vozarik2023time} for encouraging long-term liquidity provision in decentralized exchanges (DEXs), addressing challenges associated with the transient behavior of liquidity mining participants, also referred to as 'Mercenary Capital' \cite{farag2024returns, capponi2024paradox, park2023phantom}. Traditional mining incentives often lead to liquidity migrating rapidly between pools, driven by short-term opportunities to maximize annual percentage returns (APR). This behavior destabilizes liquidity pools, creating uncertainty and inefficiencies for traders \cite{heimbach2021behavior}. The time-based incentive model seeks to mitigate these issues by rewarding LPs based on the duration of their contribution, fostering stability and predictability within liquidity pools.

The mechanism employs an asymptotic reward curve, designed to favor long-term liquidity providers. Rewards are minimal during the initial days of liquidity provision, discouraging participants seeking only short-term value extraction. However, rewards increase rapidly in the subsequent weeks, achieving over $90\%$ efficiency within a two-week period and continuing to reward long-term participants relative to their loyalty. This structure ensures that while new participants can earn rewards relatively quickly, they remain at a slight disadvantage compared to longer-term providers, incentivizing extended participation.

Reward calculation is divided into two components: 1) the reward base; and 2) efficiency. The reward base is determined by the total rewards allocated to the pool, distributed among LP tokens in proportion to their contribution during a given session. Efficiency is a function of the cumulative "work" performed by the LP's capital in the pool over time, normalized against the theoretical maximum work achievable within the same period. This ensures that LPs are incentivized not only by the amount of liquidity they provide but also by the duration for which it remains active in the pool \cite{vozarik2023time}.

To achieve computational efficiency on blockchain networks, where iterating over historical data can be costly, the model discretizes the reward distribution into sessions. Each session represents a fixed time interval during which liquidity contributions are recorded, and a cumulative reward-per-liquidity metric is updated. The notion of work quantifies the productivity of capital, with cumulative work increasing asymptotically as liquidity remains active, and missed work diminishing at a predictable rate. This design enables precise and gas-efficient reward calculations while maintaining the integrity of the asymptotic reward structure.

\subsection{The GASP Token}
The GASP token, a utility token, forms the foundational layer of the Gasp Network, providing an optimized swap experience for participants. As the network progresses through its development phases, GASP will play an increasingly central role in securing the network, incentivizing participants, and ensuring efficient decentralized exchange (DEX) operations.

A core utility of GASP is facilitating gas-free swaps on the Gasp Network, achieved through Gasp's application-specific blockchain design as described herein. This eliminates the burden of gas fees, a significant cost for users on blockchains like Ethereum, where transaction costs can constitute a significant burden during periods of high network congestion. To maintain network integrity and prevent abuse from low-value or spam transactions, Gasp employs a gating mechanism requiring users to lock a predetermined amount of GASP for 24 hours for swaps under a certain value threshold. This ensures that high-volume, low-value transactions do not exploit Gasp's gasless framework, while locked tokens are released back to users after the specified period. The more GASP tokens held by a user, the greater the number of swaps they can execute within a 24-hour window, balancing cost-efficiency with robust protection against spam.

\begin{figure}[htbp]
    \centering
    \includegraphics[width=0.6\linewidth]{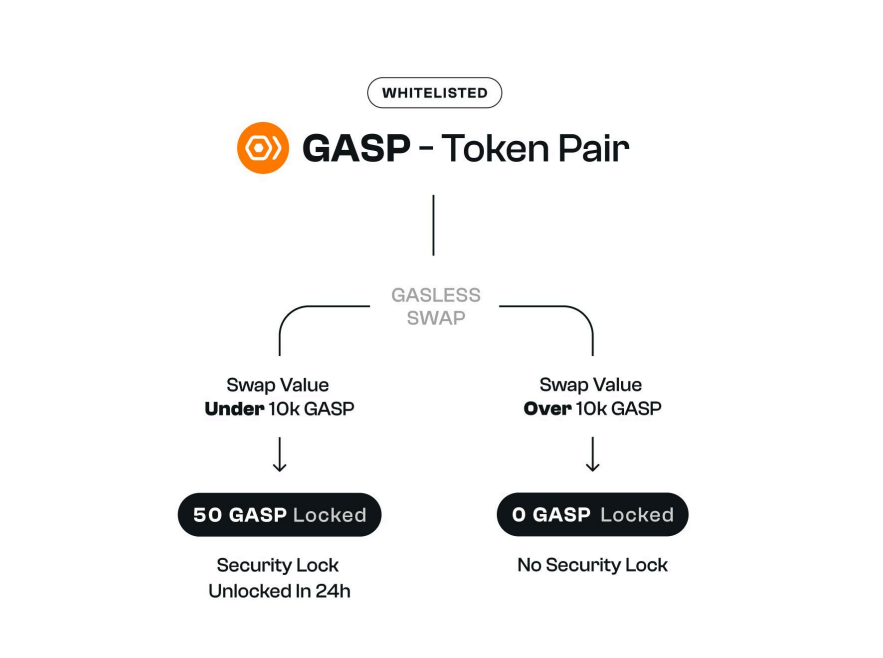}
    \caption{Gasp Token Lockup Mechanism for Network Spam Prevention.}
    \label{fig:gasp_lockup}
\end{figure}

GASP also underpins the network's security, functioning as a critical component of Gasp's status as an actively validated service (AVS) on EigenLayer. Future developments will leverage the GASP token for staking on EigenLayer, enhancing economic security and providing additional layers of trustless validation for the protocol.

Governance initiatives for the Gasp Network, built around the GASP token, will also be introduced, enabling token holders to participate in key decision-making processes. Gasp governance operates through a council of seven members tasked with proposing and voting on actions, complemented by a foundation of three members with veto authority and oversight over maintenance mode and council composition. The governance of smart contracts is similarly structured, with a seven-member council responsible for contract upgrades, appointing new council members, and managing critical updates.

A democracy mechanism, yet to be implemented, will introduce periodic voting cycles on the Gasp network, allowing any address to register as a council candidate and enabling GASP token holders to lock tokens as votes. Upon conclusion of the voting period, tokens are unlocked, and the election results are stored in the state. The transition to newly elected council members requires manual execution through a proposal created and enacted by the existing council. This ensures a deliberate and controlled governance transition.

\section{Discussion}
As stated earlier in this paper, Gasp aims to become a consolidation layer for all assets. This goal, unique to Gasp, is our view on uniting the fragmented liquidity across mutually exclusive blockchain ecosystems, not limited by the current constraints of simply swapping crypto assets and tokens, but by delivering deep liquidity to all forms of assets that can be tokenized or represented onchain from RWAs, NFTs, commodities, in-game assets, to tokenized equity, derivatives, bonds and much more. We envisage a globally connected ecosystem where any representation of value can be swapped or exchanged instantly, cheaply, and securely onchain in a decentralized way.

Gasp’s master-rollup design combines decentralized sequencing, cryptographic state verification, and deterministic finality, utilizing EigenLayer to create a robust framework for secure and scalable cross-L1 communication.

In this way, Gasp could be viewed as a restaked rollup framework or cross-chain automated market maker (AMM), where EigenLayer restakers function as node operators for L2s. It sequences and validates transactions across multiple L1s in parallel, with asset custody secured by L1 smart contracts and swap logic cryptographically validated by EigenLayer's restaked security.

Gasp’s time-based incentive system allows for seamless addition or removal of liquidity at any time without disrupting reward calculations. LPs can claim rewards or adjust their positions dynamically, with rewards recalculated based on the updated checkpoints of activity. The approach provides clear advantages for DEXs by promoting stable liquidity pools, reducing volatility, and aligning the incentives of LPs with the long-term health of the ecosystem. It also ensures fair treatment for all participants, as those contributing earlier or for longer durations are proportionately rewarded, fostering a sustainable liquidity environment.

As discussed, capital efficiency is a key metric for evaluating the performance of DeFi protocols, defined as the value generated per unit of capital deployed. Gasp’s proof of liquidity consensus mechanism maximizes Gasp’s capital efficiency by reducing transaction overhead, minimizing slippage in token swaps, and streamlining processes to reduce the number of transactions required. For Gasp, this means fewer friction points and greater value extraction from the available capital. By merging staking and liquidity provision, PoL ensures that liquidity providers (LPs) not only contribute to the platform’s liquidity but also enhance its security, aligning their incentives with the network’s broader goals.

Perhaps most importantly our solution also introduces probabilistic value extraction. In the design described above, participants will have a higher chance to claim any pure gains by sending multiple transactions, and we argue it is fair since every participant has equal opportunity. In other designs, even with Flashbots auctions, the power is shifted more toward miners, who are the ultimate decision-makers over the block content. At the same time, in Themis Consensus Extension, participants are disincentivized to attempt to extract any gains with multiple transactions, thus eliminating an edge that is making frontrunning activities statistically profitable.

An example of a common VED attack is denying pure gain transactions (e.g. arbitrage) and replacing them with their own. We argue that if only miners have this possibility, it’s an unfair design because such a pure gain opportunity is never available for a user.

There are 3 levels of VED solutions, from least to most robust:
\begin{enumerate}
    \item Miners shouldn't deny transactions,
    \item Miners don't know which transaction to deny,
    \item Miners can't deny transactions.
\end{enumerate}
The current state of the VED solution reaches the 2nd level of robustness that the Miner (or any transaction relayer) doesn't know what to deny because it has no information about the transaction purpose.

To achieve this, we propose the encryption of transactions by the user, which can be decrypted only by the transaction executor, given it's been first decrypted by the block builder. It's achieved by transactions being encrypted by public keys of both the block builder and block executor. This creates a doubly encrypted transaction that is sent to the mempool.

In the following steps, block builders are required to decrypt the transaction, but that doesn't reveal its content to them because the transaction still requires a final decryption from the block executor. In this approach, the transaction builder can't know the content of the transaction, but the executor is forced to decrypt and execute it. This means that the block builder and block executor have to be known ahead of time and submission into the mempool is not node-agnostic, but has to include that foreknowledge.

It is important to note that transaction encryption is voluntary and should be used only where it is reasonable. This is because processing an encrypted transaction is costly and the processing costs are not guaranteed to be covered by the economic value created by transaction processing. In fact, since the content of the encrypted transaction is unknown, it may even be impossible to execute. This is why, unlike unencrypted exchange transactions, all encrypted transactions are required to have fixed gas costs.

\section{Conclusion}
In conclusion, Gasp presents advancements in decentralized exchange (DEX) architecture, addressing critical challenges in cross-chain communication, capital efficiency, and user-centric design. By leveraging a master-rollup infrastructure secured by EigenLayer, Gasp achieves robust, L1-grade security while facilitating gas-free, native cross-chain swaps without the reliance on bridges or centralized intermediaries. The integration of decentralized sequencers, escape hatches, and cryptographic state transitions establishes a scalable and resilient framework that ensures asset security and user trust across diverse blockchain ecosystems.

Gasp's novel approaches, including Proof-of-Liquidity and time-based LP incentives, redefine liquidity provisioning by combining staking and liquidity provision to maximize network security and economic efficiency. These mechanisms align incentives between liquidity providers and the protocol, fostering stability and long-term engagement. Additionally, our novel application of Themis Architecture for MEV protection demonstrates a cutting-edge solution to mitigate value extraction risks, reinforcing fairness and efficiency within the platform.

By introducing the GASP utility token as a foundational element, Gasp provides a comprehensive ecosystem where security, liquidity, and governance are integrated seamlessly.

\bibliographystyle{unsrt}
\bibliography{references}

@article{augusto2022multi,
  title={Multi-Party Cross-Chain Asset Transfers},
  author={Augusto, A and Belchior, R and Hardjono, T and Vasconcelos, A and Correia, M},
  journal={TechRxiv},
  year={2022},
  publisher={TechRxiv}
}

@article{zhu2023study,
  title={A study on the challenges and solutions of blockchain interoperability},
  author={Zhu, S and Chi, C and Liu, Y},
  journal={China Communications},
  year={2023}
}

@inproceedings{augusto2024sok,
  title={SoK: Security and Privacy of Blockchain Interoperability},
  author={Augusto, A and Belchior, R and Correia, M and Vasconcelos, A and Zhang, L and Hardjono, T},
  booktitle={2024 IEEE Symposium on Security and Privacy (SP)},
  pages={3840--3865},
  year={2024}
}

@article{li2022blockchain,
  title={Blockchain Cross-Chain Bridge Security: Challenges, Solutions, and Future Outlook},
  author={Li, N and Qi, M and Xu, Z and Zhu, X and Zhou, W and Wen, S and Xiang, Y},
  journal={Proceedings of the ACM on Digital Threats: Research and Practice},
  year={2022}
}

@article{zhang2023security,
  title={Security of Cross-chain Bridges: Attack Surfaces, Defenses, and Open Problems},
  author={Zhang, M and Zhang, X and Zhang, Y and Lin, Z},
  journal={arXiv preprint arXiv:2301.13776},
  year={2023}
}

@inproceedings{lee2023sok,
  title={SoK: Not Quite Water Under the Bridge: Review of Cross-Chain Bridge Hacks},
  author={Lee, S-S and Murashkin, A and Derka, M and Gorzny, J},
  booktitle={2023 IEEE International Conference on Blockchain and Cryptocurrency (ICBC)},
  pages={1--14},
  year={2023}
}

@misc{vozarik2023cross,
  title={Cross-L1 information transfer via master rollup secured by EigenLayer},
  author={Vozarik, S and Urvanov, G and Siddiqui, S and Kris, P},
  howpublished={Eigen Layer Forum},
  year={2023}
}

@article{li2023decentralized,
  title={Decentralized Sequencers in Rollup Architectures},
  author={Li, Y and Zhang, X and Wang, H},
  journal={arXiv preprint arXiv:2310.03616},
  year={2023}
}

@misc{vozarik2023themis,
  title={A Solution to MEV for Application-specific Blockchains},
  author={Vozarik, S and Siddiqui, S and Kris, P and Urvanov, G},
  howpublished={Gasp Docs},
  year={2023}
}

@article{bartoletti2023theoretical,
  title={A Theoretical Basis for Blockchain Extractable Value},
  author={Bartoletti, M and Zunino, R},
  journal={arXiv preprint arXiv:2302.02154},
  year={2023}
}

@article{braga2024redistribution,
  title={On the Redistribution of Maximal Extractable Value: A Dynamic Mechanism},
  author={Braga, P and Chionas, G and Leonardos, S and Krysta, P and Piliouras, G and Ventre, C},
  journal={arXiv preprint arXiv:2402.15849},
  year={2024}
}

@misc{vozarik2023proof,
  title={Proof of Liquidity},
  author={Vozarik, S and Siddiqui, S and Kris, P and Urvanov, G},
  howpublished={Gasp Docs},
  year={2023}
}

@misc{vozarik2023time,
  title={Time-Based LP Incentives},
  author={Vozarik, S and Siddiqui, S+ and Kris, P and Urvanov, G},
  howpublished={Gasp Docs},
  year={2023}
}

@article{farag2024returns,
  title={Returns from Liquidity Provision in Cryptocurrency Markets},
  author={Farag, H and Luo, D and Yarovaya, L and Zi{\k{e}}ba, D},
  journal={SSRN Electronic Journal},
  year={2024}
}

@article{capponi2024paradox,
  title={The Paradox of Just-in-Time Liquidity in Decentralized Exchanges: More Providers Can Lead to Less Liquidity},
  author={Capponi, A and Jia, R and Zhu, B},
  journal={SSRN Electronic Journal},
  year={2024}
}

@article{park2023phantom,
  title={Phantom Liquidity in Decentralized Lending},
  author={Park, A and Stinner, J},
  journal={SSRN Electronic Journal},
  year={2023}
}

@article{heimbach2021behavior,
  title={Behavior of Liquidity Providers in Decentralized Exchanges},
  author={Heimbach, L and Wang, Y and Wattenhofer, R},
  journal={arXiv preprint arXiv:2105.13822},
  year={2021}
}

\end{document}